\documentclass[twocolumn]{revtex4}

\usepackage{graphicx}

\begin{document}

\title{Enhancement of electron spin coherence
       by optical preparation of nuclear spins}

\author{Dimitrije Stepanenko}
\author{Guido Burkard}
\affiliation{Department of Physics and Astronomy,
             University of Basel,
             Klingelbergstrasse 82,
             CH-4056 Basel, Switzerland}

\author{Geza Giedke}
\author{Atac Imamoglu}
\affiliation{Institute of Quantum Electronics,
             ETH Z\"urich,
             CH-8093 Z\"urich, Switzerland}

\begin{abstract}
We study a large ensemble of nuclear spins 
interacting with a single electron spin in a quantum dot under optical
excitation and photon detection.
When a pair of applied laser fields satisfy two-photon resonance between the two ground 
electronic spin states, detection of light scattering from the intermediate
exciton state acts as a weak quantum measurement of the effective magnetic
(Overhauser) field due to the nuclear spins.
If the spin were driven into a coherent 
population trapping state where no light scattering takes place, then the nuclear state 
would be projected into an eigenstate of the Overhauser field operator
and electron decoherence due to nuclear spins would be suppressed: 
we show that this limit can be approached by
adapting the laser frequencies when a photon is detected.
We use a Lindblad equation to describe the time evolution of the
driven system under photon emission and detection.
Numerically, we find an increase of the electron coherence
time from $5\,{\rm ns}$ to $500\,{\rm ns}$ after
a preparation time of 10 microseconds.
\end{abstract}


\maketitle

\textit{Introduction.}
Single electron spins localized in small artificial
structures, such as semiconductor quantum dots (QDs),
have become available and to a large extent controllable
\cite{Elzerman,Johnson,Koppens,Petta}.  Of particular interest is the
phase coherence of such electron spins as single quantum
objects, both from a fundamental physics point of view and because of
their potential use as quantum bits (qubits) for
quantum information processing \cite{LD98,Imamoglu99}.

A number of physical mechanisms that lead to the gradual reduction of
the quantum phase coherence (decoherence) of the electron spin have
been analyzed \cite{Cerletti}.   It has been established
experimentally and theoretically that for an electron in a GaAs QD, 
the predominant decoherence mechanism is the hyperfine coupling to the 
nuclear spins in the host material \cite{BLD99,KLG02,MER02,KLG03,CL04,CL05}.  
For an unpolarized ensemble of $N$ nuclei and an effective hyperfine 
interaction energy $A$, the dephasing time in the presence
of a weak magnetic field is $T_2^*\sim 1/\sigma \sim \sqrt{N}/A$ where
$\sigma$ is the width of the distribution of nuclear field values
$h_z$ parallel to the field.  
In a typical GaAs QD with 
$A\sim 90\,\mu{\rm eV}$ or $A/g\mu_B=3.5\,{\rm T}$ \cite{Paget}, 
the number of Ga and As nuclei (spin $I=3/2$) is $N\sim 5\cdot 10^5$ and
$T_2^*\sim 5\,{\rm ns}$; this value is supported by the
experimental evidence \cite{Petta,footnote1}.
The $T_2^*$ decay originates from nuclear ensemble
averaging and can be prolonged by narrowing the nuclear spin
distribution \cite{CL04}.  Another strategy 
is to polarize the nuclear spins \cite{BLD99}, but
this requires a polarization close to 100\%
which is currently not available \cite{CL04}.
Two schemes have been proposed to achieve a narrowing of the nuclear
spin distribution, based on electron transport \cite{Giedke} and
gate-controlled electronic Rabi oscillations \cite{KCL}.

\begin{figure}[t]
 \centerline{\includegraphics[width=8cm]{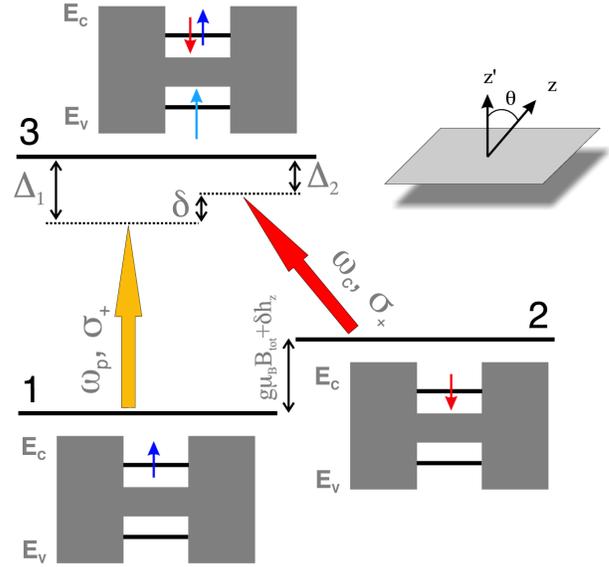}}
\caption{
Three-level system.  State $1$ ($2$) 
is a spin-up (down) conduction-band ($E_C$) electron,
with splitting $g\mu_B B_{\rm tot}+\delta h_z$ 
where $\delta h_z$ is the $z$-component
of the nuclear (Overhauser) field fluctuations.
State $3$ is a trion with $J_{z'}=3/2$.
Two laser fields with frequencies 
$\omega_p$ and $\omega_c$ are applied near 
the $13$ and $23$ resonances with detunings $\Delta_{1,2}$.  
For $\sigma_+$ circularly polarized excitation (along $z'$),
both transitions are allowed for $\theta\neq 0$ and transitions 
to the $J_{z'}=-3/2$ states are forbidden.
Inset:  Structural axis $z'$, leading to a splitting in $E_V$
and spin quantization axis $z\parallel {\bf B}_{\rm tot}$ in $E_C$
where $\cos \theta=z\cdot z'<1$.
\label{fig:levels}}
\end{figure}
Here, we analyze an optical scheme for nuclear spin preparation
that makes use of spin-flip two-photon (Raman) resonance in a 
driven three-level system (TLS), in analogy to
electromagnetically induced transparency (EIT) in atomic
systems \cite{EIT}.  The lowest electronic states in a QD 
formed in a III-V semiconductor (e.g., GaAs) that are
optically active under $\sigma_+$ circularly polarized
excitation are a Zeeman-split single electron in the localized 
conduction band ($E_C$) ground state and the negatively
charged exciton (trion) $|X\rangle$, i.e., two 
electrons (spin up and down) plus one valence
band heavy hole (hh) with angular momentum $J_{z'}=+3/2$
(Fig.~\ref{fig:levels}).
The $J=3/2$ sector in the valence band is split into
lh and hh states along the quantization axis ${z'}$
determined by the direction of strong QD confinement.
Here, we assume
excitation from the hh ($J_{z'}=\pm 3/2$) subband only.
The quantization axis $z$ in $E_C$ is parallel to
the total magnetic field ${\bf B}_{\rm tot}$, and we assume that
the axes $z$ and ${z'}$ enclose an angle $\theta > 0$. The spin
up and down states in $E_C$ are then
$|\!\!\uparrow\rangle \equiv 
|\!\!\uparrow\rangle_z = \cos(\theta)|\!\!\uparrow\rangle_{z'} +
\sin(\theta)|\!\!\downarrow\rangle_{z'}$
and 
$|\!\!\downarrow\rangle \equiv
|\!\!\downarrow\rangle_z =
\cos(\theta)|\!\!\downarrow\rangle_{z'} -\sin(\theta)|\!\!\uparrow\rangle_{z'}$.
  From now on, we drop the index $z$. 
Two circularly polarized ($\sigma_+$) continuous-wave lasers 
at the frequencies $\omega_p=\omega_X-\omega_\uparrow-\Delta_1$ and
$\omega_c=\omega_X-\omega_\downarrow-\Delta_2$ stimulate the
transitions between $|\!\!\uparrow\rangle$ and $|X\rangle$ and
between $|\!\!\downarrow\rangle$ and $|X\rangle$, while the trion
with $J_{z'}=-3/2$ is not excited.

The mechanism leading to a narrowing of the nuclear 
field distribution $\nu$ works as follows.  
The population of the excited state $|X\rangle$, and thus
the probability of photon emission (scattering) is only
non-zero away from the two-photon resonance $\delta=0$ where
$\delta = \Delta_1 -\Delta_2$.
In the presence of the nuclear spins this resonance
moves to $\delta=\delta h_z$ where $\delta h_z$ is the
deviation of the nuclear field (along $z$)
from its mean $\langle h_z \rangle$.
The absence of photon emission during
a waiting time $t$ constitutes a weak measurement
of the quantum operator $\delta h_z$.  In the limit
$t\rightarrow\infty$, it becomes
a strong measurement, projecting the nuclear state
onto $|\delta h_z=0\rangle$ (width $\sigma=0$),
thus eliminating electron decoherence 
due to the fluctuating field $\delta h_z$.

\textit{Model.}
The Hamiltonian for the TLS coupled to nuclei,
\begin{equation}
  \label{eq:1}
  H = H_0 + H_{\rm int} + H_{\rm hf},
\end{equation}
contains the three energy levels in
$H_0 = -\frac{\hbar \omega_z}{2}  \Sigma_z + \hbar\omega_X P_X$,
with  $\Sigma_i = \left(\begin{array}{c c} \sigma_i & 0 \\ 0 &
0\end{array}\right)$ and $P_X=|X\rangle\langle X|=(0 0 1)^T(0 0 1)$.
The spin splitting is given as
$\hbar\omega_z 
= g\mu_B B_{\rm tot}
= \left|g\mu_B {\bf B} + \langle {\bf h} \rangle\right|$,
the sum of the external magnetic and the mean nuclear fields.
The nuclear (Overhauser) field operator is
${\bf h} = \sum_{i=1}^N A_i {\bf I}_i$,
where $A_i = a_i v_0 |\Psi({\bf r}_i)|^2$, and $\Psi({\bf r}_i)$
denotes the electron wave function at the position ${\bf r}_i$ of the
$i$th atomic nucleus and $v_0$ is the volume of the unit cell.
The classical laser
fields in the rotating wave approximation (RWA) are described by \cite{EIT}
$H_{\rm int} = \Omega_p e^{i\omega_p t}|X\rangle\langle  \uparrow\!\!|
+\Omega_c e^{i\omega_c t}|X\rangle\langle  \downarrow\!\!|+{\rm h.c.}$
The coupling of the electron spin to the quantum fluctuations of the
Overhauser field ${\bf h}$, will be described
by the term 
$H_{\rm hf} = -\frac{1}{2} \delta{\bf h}\cdot {\bf \Sigma}$,
where $\delta {\bf h} = {\bf h} - \langle{\bf h}\rangle$.
In the rotating frame
$\tilde\Psi (t) = U(t)\Psi (t)$
with $U(t) = e^{-i\omega_p t} P_\uparrow +  e^{-i\omega_c t} P_\downarrow+P_X$,
where $P_{\uparrow} = |\!\!\!\uparrow\rangle\langle\uparrow\!\!\!|$
and $P_{\downarrow} = |\!\!\!\downarrow\rangle\langle\downarrow\!\!\!|$, 
we find
$\tilde H(t) = U(t)\left[H(t)+\hbar\omega_p P_\uparrow 
+\hbar\omega_c P_\downarrow\right]U(t)^\dagger$, and, 
up to constant terms proportional to the unity matrix
(we drop the tilde and use $H$ for the Hamiltonian henceforth),
\begin{equation}
    H(t) = -\frac{\hbar}{2}\left(\begin{array}{c c c} \delta   &
      0        & \Omega_p \\ 0        & -\delta  & \Omega_c \\
      \Omega_p & \Omega_c & -\Delta
\end{array}\right)
 - \frac{\hbar}{2}\delta h_z \Sigma_z + H_{\perp}   \label{eq:11}
\end{equation}
where $\Delta = \Delta_1+\Delta_2$.  
The hyperfine flip-flop terms 
$H_{\perp} = \hbar\left(\delta h_+ \Sigma_- e^{it(\omega_p-\omega_c)}
+ \delta h_- \Sigma_+ e^{-it(\omega_p-\omega_c)}\right)/4$
are oscillating rapidly at the frequency $\omega_p-\omega_c =
g\mu_B B_{\rm tot}/\hbar -\delta$ and can be 
neglected in the RWA \cite{footnote2}, leading to a block-diagonal
Hamiltonian $H={\rm diag}(H_1, H_2, \ldots, H_K)$, with 
\begin{equation}
  \label{eq:13}
    H_k = -\frac{\hbar}{2}\left(\begin{array}{c c c} \delta h_z^k +
      \delta    & 0                    & \Omega_p \\ 0 & -\delta
      h_z^k-\delta & \Omega_c \\ \Omega_p                 & \Omega_c &
      -\Delta
\end{array}\right),
\end{equation}
where $\delta h_z^k$ are the
eigenvalues of the operator $\delta h_z$.
The state of the TLS combined with the nuclear spins is described
by the density matrix $\rho$, which we divide up into 3-by-3 
blocks $\rho_{kk'}$.
The density matrix evolves according to the generalized master
equation \cite{EIT}
\begin{equation}
  \label{eq:17}
  \dot\rho = {\cal L}\rho \equiv
  \frac{1}{i\hbar}\left[H,\rho\right]+{\cal W}\rho,
\end{equation}
with the Hamiltonian  Eq.~(\ref{eq:11}) and the dissipative term
${\cal W}\rho = \sum_{\alpha=\uparrow,\downarrow} 
\Gamma_{X\alpha} (2\sigma_{\alpha X}\rho\sigma_{X \alpha} 
- \sigma_{XX}\rho - \rho\sigma_{XX})/2 
+\sum_{\beta=\downarrow,X} \gamma_{\beta} 
(2\sigma_{\beta\beta}\rho\sigma_{\beta\beta} -
\sigma_{\beta\beta}\rho - \rho\sigma_{\beta\beta})/2$,
where $\sigma_{ij}=\sigma_{ij}\otimes\openone=|i\rangle\langle j|$
acts on the TLS only.  The rate $\Gamma_{X\alpha}$
describes the radiative decay of the exciton $|X\rangle$ into one of the
single-electron states $\alpha=|\!\!\uparrow\rangle,|\!\downarrow\rangle$,
while $\gamma_\beta$ is the pure
dephasing rate of state $\beta=|\!\downarrow\rangle, |X\rangle$ 
with respect to $|\!\!\uparrow\rangle$.
Since $H$ has block form, the master equations for the
various blocks are decoupled, and have the closed form
\begin{equation}
  \dot\rho_{kk'} =
  \frac{1}{i\hbar}\left(H_k\rho_{kk'}-\rho_{kk'}H_{k'}\right)   +
  {\cal W}\rho_{kk'}.
\label{eq:18}
\end{equation}
The diagonal blocks obey the familiar Lindblad equation,
\begin{equation}
  \label{eq:18b}
  \dot\rho_{kk} = {\cal L}_k \rho_{kk},
\quad\quad {\cal L}_k = -i[H_k,\rho]+{\cal W}\rho.
\end{equation}

\begin{figure}
\centerline{\includegraphics[height=8.6cm,angle=270]{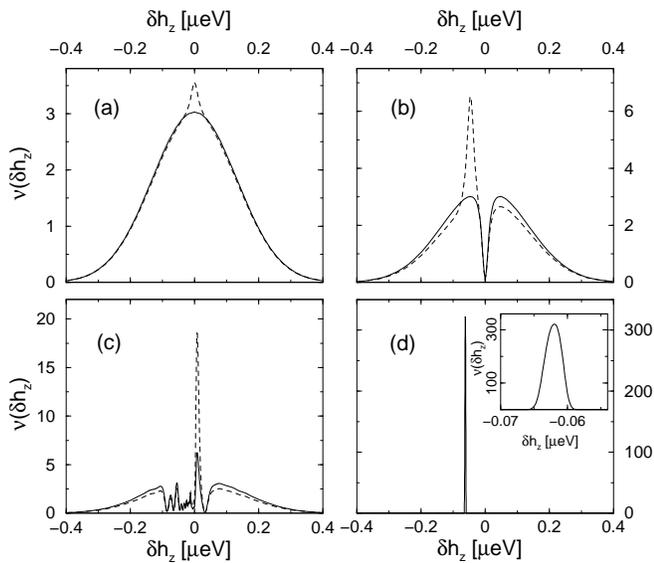}}
\caption{Conditional evolution of the nuclear spin
  distribution $\nu(\delta h_z^k)=\nu_{kk}$.  (a) During the
  first period $t_1$ of evolution without photon emission, the initial
  Gaussian distribution (solid line) develops a
  peak at the two-photon resonance (dashed line).
  (b) Change of $\nu(\delta h_z)$
  after emission at $t_1$ (solid line), until
  before emission time $t_2$ of the second photon (dashed line).
  The two-photon
  resonance $\delta$ has been shifted to the position of the left maximum 
  (adaptive technique).
  The depleted region around $\delta h_z^k=0$ is developed at $t_1$.
  (c) Analogous situation between $t_{11}$ and $t_{12}$.
  (d) $\nu(\delta h_z)$ is obtained after a total time of $10 \mu {\rm s}$.
  Inset: Magnification of peak in (d). The width of
  $\nu(\delta h_z)$ is reduced by a factor of
  $\approx 100$ compared to the initial width in (a).
  The parameters used for this calculation are
  $\Omega_c=\Omega_p=0.2\,{\rm ns}^{-1}$, $\Delta=0$,
  $\Gamma_{X\uparrow}=\Gamma_{X\downarrow}=1\,{\rm ns}^{-1}$,
  and $\gamma_{\downarrow}=\gamma_{X}=0.001\,{\rm ns}^{-1}$.
\label{fig:first}}
\end{figure}

\textit{Stationary state.}
We start with the factorized initial state 
$\rho_0  = \chi_0 \otimes \nu_0$
with arbitrary initial density matrices $\chi_0$ and $\nu_0$ 
of the TLS and the nuclear ensemble, where
$\nu_0 = \sum_{kk'} \nu_{kk'} |\delta h_z^k \rangle \langle\delta h_z^{k'}|$,
and $|\delta h_z^k\rangle$ are eigenstates of the Overhauser
operator, $\delta h_z|\delta h_z^k\rangle = \delta
h_z^k|\delta h_z^k\rangle$.  
As the off-diagonal elements of $\nu_0$ turn out to be
irrelevant for the stationary state of the TLS,
our analysis is valid both for a pure initial state
(e.g., with $\nu_{kk'}=\sqrt{\nu_{kk}\nu_{k'k'}}$) and for
a mixed initial state (e.g., the completely mixed state,
$\nu_{kk'}\propto \delta_{kk'}$).
We assume a Gaussian distribution  $\nu_{kk} =
(2\pi)^{-1/2}\sigma^{-1}\exp[-(\delta h_z^k)^2/2\sigma^2]$, with an
initial width $\sigma=\sigma_0=A/\sqrt{N}$,
plotted as a solid line in Fig.~\ref{fig:first}(a).  For our numerical
calculations we choose $A=90\,\mu{\rm eV}$ and $N\approx 5\cdot 10^5$
corresponding to $\sigma_0\simeq 0.13\,\mu{\rm eV}\simeq
0.2\,\hbar\Gamma$ with $\Gamma=1\,{\rm ns}$.
We choose a sample of $n\ll (2I+1)^N$ states from the total Hilbert
space ($n\sim 4000$).

Due to the hyperfine coupling between the TLS and the nuclei,
the two systems become entangled as the stationary
state is reached.  The stationary density
matrix $\bar\rho$ is independent of $\chi_0$ and has the general form
$\bar\rho= \sum_{kk'}\bar\rho_{kk'} \otimes |\delta
h_z^k\rangle\langle \delta h_z^{k'}|$.
We derived an analytical expression for
the 3-by-3 diagonal blocks $\bar\rho_{kk}$ of $\bar\rho$
as a function of all parameters, including $\delta h_k$.  
We find numerically that the off-diagonal blocks $\rho_{kk'}$ ($k\neq
k'$), i.e., the nuclear coherences, decay exponentially and thus vanish in
the stationary density matrix $\bar\rho$. Therefore,
$\bar\rho$ is block-diagonal, $\bar\rho_{kk'} =
\delta_{kk'}\bar\rho_{kk}$.

\textit{Evolution of the observed system.}
In order to enhance the electron spin coherence, we aim at
\textit{narrowing}  the nuclear spin distribution.  In the case of
a Gaussian distribution, this amounts to decreasing the width
$\sigma$ with respect to its initial value $\sigma_0$, thus
increasing the electron coherence time $t_0\simeq 1/2\sigma$.
Ideally, we would perform a projective measurement $P$ on the
nuclear spins such that 
$P\bar\rho_{kk} P \propto \delta(\delta h_z^k-\delta)$.  
This also bounds the off-diagonal elements of 
$\nu={\rm Tr}_\Lambda \rho$, where ${\rm Tr}_\Lambda$ is the partial trace
over the TLS, because 
$|\nu_{kk'}|\le \sqrt{\nu_{kk}\nu_{k'k'}}$ 
due to the positivity of $\nu$.  
The measurement $P$ can be successively approximated by
monitoring the photon emission from the QD.  The
longer the period $t$ during which no photon is emitted, 
the higher is the probability for $\delta h_z$ to be at the two-photon
resonance, $\delta h_z=\delta$.

To describe the state of the system conditional on a certain
measurement record, we use the \textit{conditional density matrix} $\rho^c$.
In the absence of photon emission, $\rho_c$ obeys Eq.~(\ref{eq:18b})
with ${\cal L}_k$ replaced by ${\cal L}_k-{\cal S}$,
with the collapse operator ${\cal S}$ \cite{Carmichael},
\begin{equation}
  \label{eq:30}
  \dot\rho^c_{kk} = \left({\cal L}_k - {\cal S}\right) \rho^c_{kk},
\quad \:
   {\cal S}\rho =  \sum_{\alpha=\uparrow,\downarrow}
       \Gamma_{X\alpha} \sigma_{\alpha X}\rho\sigma_{X \alpha}.
\end{equation}
We have numerically calculated $\rho^c$ conditional on the absence
of emitted photons for a given duration $t$.  
We plot the updated probability distribution $\nu_{kk}$ from
$\nu = {\rm Tr}_\Lambda \rho^c$ as a dashed line in
Fig.~\ref{fig:first}(a).  
We find that the \textit{a posteriori}
probability is concentrated around the two-photon resonance.
Of course, this process will eventually be stopped by the emission
of a photon.
\begin{figure}[t]
\centerline{\includegraphics[height=7.7cm,angle=270]{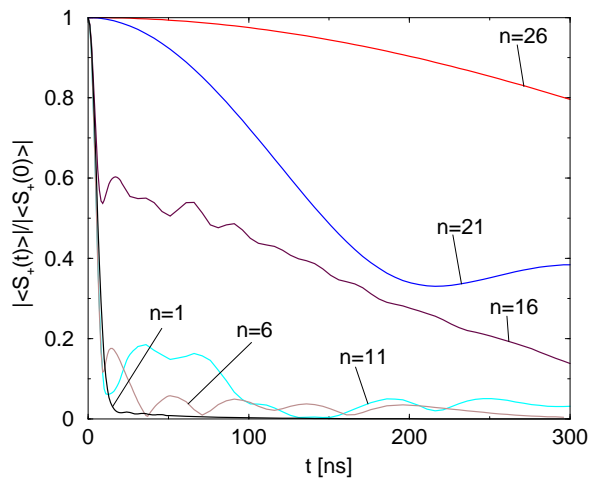}}
\caption{Electron coherence function
  $|\langle S_+(t)\rangle|/|\langle S_+(0)\rangle|$
  vs.\ electronic precession time $t$ calculated from
  $\nu(\delta h_z)$ in Fig.~\ref{fig:first}
  after emission of the $n$th photon ($n=1, 6, \ldots, 26$).
  The initial decay is approximately Gaussian.
\label{fig:splus}}
\end{figure}

\textit{Photon emission.}
The stationary emission rate is \cite{Carmichael}
\begin{equation}
  \label{eq:32}
  \Gamma_{\rm em} = {\rm Tr} {\cal S}\bar\rho(t)
     =  \Gamma \sum_k(\rho_{kk})_{XX} \nu_{kk},
\end{equation}
where $\Gamma =\Gamma_{X\uparrow}+\Gamma_{X\downarrow}$.
The average number of photons emitted during time $t$ is
$\langle N_{\rm ph}\rangle = t \Gamma_{\rm em}$
and the \textit{a priori} probability for $N_{\rm ph}=0$
is, according to Poissonian statistics,
$P_{\rm dark}(t) = \exp(-\Gamma_{\rm em} t)$.
The waiting time distribution for photon emissions is
$p_{\rm wait}(t) = \Gamma_{\rm em}^{-1}\exp(-\Gamma_{\rm em} t)$
with a mean waiting time $\langle t\rangle = \Gamma_{\rm em}^{-1}$.
The progressive narrowing of the Overhauser field distribution,
Eqs.~(\ref{eq:30}) and (\ref{eq:32}),
leads to a decreasing photon emission rate $\Gamma_{\rm em}$,  
and therefore to an increasing average waiting time $t$.
We first assume that every emitted photon is detected and later
generalize to the case of imperfect detection.

Using Eq.~(\ref{eq:32}), we find for the update rule of
the nuclear density matrix upon photon emission
$\nu' = {\rm Tr}_\Lambda {\cal S}\rho^c/{\rm Tr}{\cal S}\rho^c$.
The Overhauser field distribution after the emission is  
\begin{equation}
\label{eq:nu_diagonal}
 \nu_{kk}'
 =\frac{\nu_{kk}(\rho_{kk})_{XX}}{\sum_j\nu_{jj}(\rho_{jj})_{XX}},
\end{equation}
where $\nu_{kk}$ and $(\rho_{kk})_{XX} =\langle X|\rho_{kk}| X\rangle$ are 
taken before the emission. 
According to Eq.~(\ref{eq:nu_diagonal}), the
population in the Overhauser field $\delta h_z$ corresponding to the
two photon resonance $\delta h_z=\delta$ is depleted by the
photon emission (Fig.~\ref{fig:first}b, solid line).
However, the combined effect of the observed evolution
without emission and the collapse leads to narrowing of the nuclear
distribution.

\textit{Adaptive technique.}
The stationary, isolated TLS at the
two-photon resonance is in a dark state.  
However, the coupling to the nuclei introduces a nonzero probability 
for occupation of the $|X\rangle$ state and for emission of a photon.  
The detection of a photon provides information about 
$\delta h_z$ and an adjustment of the EIT setup. 
Thus, photon emission does not necessarily
signify a failed attempt to narrow the nuclear field distribution, 
but can be used as an input for the next weak measurement
with adjusted frequencies of the driving lasers, 
$\omega_p'=\omega_p+\epsilon/2$ and $
\omega_c'=\omega_c-\epsilon/2$, so that the new two-photon
resonance condition is $\delta h_z=\delta'$ where
$\delta'=\delta+\epsilon$ while $\Delta'=\Delta$. 
We choose $\epsilon$ such that the new
resonance with the Overhauser field lies in one of the two maxima
$\delta h_z^{\rm max}$ 
formed after the photon emission, see Fig.~\ref{fig:first}b.  This
situation is formally described by Eq.~(\ref{eq:13}) with the
substitution $\delta \rightarrow \delta + \delta h_z^{\rm max}$.
We note that the adaptive technique also works by changing only
one of the laser frequencies.
Right after the photon emission, the TLS is in one of the single
electron states, $|\!\!\uparrow\rangle$ or $|\!\!\downarrow\rangle$.  
Within a time $1/\Gamma$, much faster than any nuclear time scale,
the driven electronic system will reach the new stationary state.
Then, the photon emission from the QD 
can again be monitored, leading to an enhanced nuclear population at the
new resonance (Fig.~\ref{fig:first}b, dashed line), thus further
narrowing the nuclear distribution and further enhancing the
electron spin coherence.  This procedure can be repeated many
times, leading to a nuclear distribution that is limited only by the
width of the EIT resonance (Fig.~\ref{fig:first}c,d).
\begin{figure}[t]
\centerline{\includegraphics[width=7.7cm]{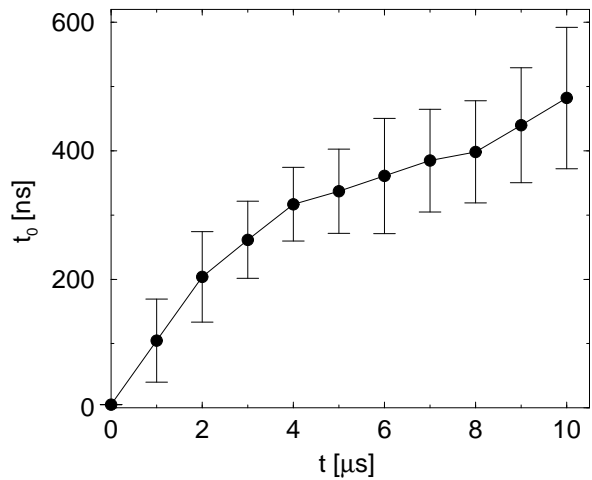}}
\caption{
  Characteristic time $t_0$ of the initial Gaussian decay in
  $|\langle S_+(t)\rangle|/|\langle S_+(0)\rangle|$ in Fig.~\ref{fig:splus}
  as a function of the optical preparation time $t$,
  averaged over 50 numerical runs (with error bars indicating the
  standard deviation).
\label{fig:coherence}}
\end{figure}

\textit{Electron spin decoherence.}
The electron spin coherence is quantified using the
expectation value of the raising operator $S_+(t)$ in a
state $|x_+\rangle$ that is prepared perpendicular to the total field
${\bf B}_{\rm tot}$ and is freely precessing about the fluctuating
nuclear field $\delta h_z$,
$\langle S_+(t)\rangle \equiv \langle x_+|S_+(t)|x_+\rangle
= \sum_k \nu_{kk} \langle \delta h_z^k|\langle x_+| 
S_+(t)|x_+\rangle|\delta h_z^k\rangle $.
Using $\langle \delta h_z^k|S_+(t)|\delta h_z^k\rangle 
= e^{it\delta h_z^k\sigma_z/2} S_+(0)e^{-it\delta h_z^k\sigma_z/2}$, 
we obtain
$\langle S_+(t)\rangle = (\hbar/2)\sum_k \nu_{kk} \exp(it\delta h_z^k)$,
which we plot in Fig.~\ref{fig:splus}
at various stages in an adaptive optical measurement scheme.
We make a Gaussian fit,
$\langle S_+(t)\rangle  \propto \exp(-t^2/t_0^2)$,
for short times $t$
and plot the coherence time $t_0$ as a function of total
waiting time in Fig.~\ref{fig:coherence}.
This is the main result of our theoretical analysis:
The repeated observation of the QD photon emission
and adaptation of the laser frequencies $\omega_c$ and $\omega_p$
after each photon emission leads
to a pronounced enhancement of the electron coherence time, for the realistic
parameters chosen, from $t_0=5\,{\rm ns}$ to $\approx 500\,{\rm ns}$
within a total observation time of $10\,\mu{\rm s}$.

\textit{Imperfect detectors.}
We cannot expect to have perfect photon detectors at our disposal,
therefore we discuss here the case of a detector with efficiency $e<1$.
For an imperfect detector, Eq.~(\ref{eq:30}) becomes
$\dot\rho^c_{kk} = \left({\cal L}_k - e {\cal S}\right) \rho^c_{kk}$,
reflecting that a photon is only detected with probability $e$.
We have numerically analyzed the case of $e=10\%$ (other parameters as above) 
and find $t_0\approx 460\,{\rm ns}$ after a preparation time of $t=50\,\mu{\rm s}$.
The requirement for longer preparation times is a consequence of the reduced
photon detection rate $e\Gamma_{\rm em}$.
We note that this is still orders of magnitude shorter than the time after 
which the nuclear spin diffuses, estimated to be around $0.01\,{\rm s}$ 
due to higher-order hyperfine flip-flop terms \cite{KCL}, but possibly
longer due to Knight shift gradient effects.  Long time scales for the
decay of nuclear polarization, on the order of seconds in the case of nuclear
spins in contact with donors in GaAs, have also been seen experimentally \cite{Paget82}.

\textit{Conclusions.}
We find that it is possible to efficiently enhance the
quantum phase  coherence of an electron spin in a QD surrounded
by a large ensemble of nuclear spins by a continuous weak measurement of the
Overhauser field using optical excitation at a two-photon resonance of the
TLS formed by the spin-split conduction band electron and one of the trion states. 
An intriguing question is whether the electron spin coherence can be enhanced by a 
quantum Zeno type effect to the point where it is ultimately determined by spin-orbit 
interaction: since reservoir correlation time of dominant electron spin decoherence 
due to flip-flop terms of the hyperfine interaction is $\sim1 \mu {\rm s}$, this would most 
likely require high efficiency detection of the scattered photons.

\textit{Acknowledgments.}
We thank W. A. Coish, D. Klauser, and D. Loss for useful discussions.
We acknowledge financial support from the Swiss National Science Foundation (SNF)
through an SNF professorship (G.B.) and through NCCR Nanoscience.


\begin{thebibliography}{99}

\bibitem{Elzerman}
J. M. Elzerman \textit{et al.},
Phys.\ Rev.\ B {\bf 67}, 161308 (2003).

\bibitem{Johnson}
A. C. Johnson \textit{et al.},
Nature {\bf 435}, 925 (2005).

\bibitem{Koppens}
F. H. L. Koppens \textit{et al.},
Science {\bf 309}, 1346 (2005).

\bibitem{Petta} 
J. R. Petta \textit{et al.},
Science {\bf 309}, 2180 (2005).

\bibitem{footnote1}
The decay of a spin echo 
envelope, being a measure for $T_2 \ge T_2^*$, can be much slower \cite{Petta}.

\bibitem{LD98}
D. Loss and D. P. DiVincenzo, Phys.\ Rev.\ A {\bf 57}, 120 (1998).

\bibitem{Imamoglu99} 
A. Imamoglu \textit{et al.},
Phys.\ Rev.\ Lett.\ {\bf 83}, 4204 (1999).

\bibitem{Cerletti} 
V. Cerletti, W. A. Coish, O. Gywat, and D. Loss,
Nanotechnology {\bf 16}, 27 (2005).

\bibitem{BLD99} G. Burkard, D. Loss, and D. P. DiVincenzo,
Phys.\ Rev.\ B {\bf 59}, 2070 (1999).

\bibitem{KLG02}
A. Khaetskii, D. Loss, and L. Glazman, 
Phys.\ Rev.\ Lett.\ {\bf 88}, 186802 (2002).

\bibitem{MER02} 
I. A. Merkulov, A. L. Efros, and M. Rosen,
Phys.\ Rev.\ B {\bf 65}, 205309 (2002).

\bibitem{KLG03} 
A. Khaetskii, D. Loss, and L. Glazman,
Phys.\ Rev.\ B {\bf 67}, 195329 (2003).

\bibitem{CL04} 
W. A. Coish and D. Loss,
Phys.\ Rev.\ B {\bf 70}, 195340 (2004).

\bibitem{CL05} 
W. A. Coish and D. Loss, 
Phys.\ Rev.\ B {\bf 72}, 125337 (2005).

\bibitem{Paget} 
D. Paget, G. Lampel, B. Sapoval, and V. I. Safarov,
Phys.\ Rev.\ B {\bf 15}, 5780 (1977).

\bibitem{Giedke} 
G. Giedke, J. M. Taylor, D. D'Alessandro,
M. D. Lukin, and A. Imamo\u{g}lu, quant-ph/0508144.

\bibitem{KCL} D. Klauser, W. A. Coish, and D. Loss, cond-mat/0510177.

\bibitem{EIT} M. Fleischhauer, A. Imamoglu, and J. P. Marangos, 
Rev.\ Mod.\ Phys.\ {\bf 77}, 633 (2005).

\bibitem{footnote2}
The neglected flip-flop terms contribute to electron-spin decoherence due to higher order terms.

\bibitem{Carmichael} 
H. J. Carmichael, 
\textit{An Open Systems Approach to Quantum Optics} 
(Springer, Berlin, 1993).

\bibitem{Paget82}
D. Paget, Phys.\ Rev.\ B {\bf 25}, 4444 (1982).


\end{thebibliography}
\end{document}